\documentclass[review]{elsarticle}

\usepackage{hyperref}

\usepackage{graphicx}
\usepackage{dcolumn}
\usepackage{newlfont,epsfig}
\usepackage{graphicx}
\usepackage{epstopdf}
\usepackage{amsmath}

\usepackage{color}
\usepackage{amsmath,mathrsfs}
\usepackage{verbatim}
\usepackage{multirow}
  \usepackage[thinlines]{easytable}
  \usepackage{braket}
  \usepackage[varg]{txfonts}

  \usepackage{newlfont}
  \usepackage{amssymb}
  \usepackage[latin1]{inputenc}
  \usepackage{bbm}

\journal{Physics Letters A}

\begin{document}

\begin{frontmatter}

\title{Quantum resources of the steady-state of three coupled qubits: Microscopic versus Phenomenological model}

\author{E. C. Diniz$^1$, A. C. S. Costa$^2$, and L. K. Castelano$^1$}
\address{$^1$Departamento de F\'{i}sica, Universidade Federal de S\~{a}o Carlos, P.O. Box 676, 13565-905, S\~{a}o Carlos, S\~{a}o Paulo, Brazil. \\
$^2$Departamento de F\'{i}sica, Universidade Federal do Paran\'{a}, Caixa Postal 19044, 81531-980, Curitiba, Paran\'{a}, Brazil.}


\begin{abstract}
Quantum resources, such as entanglement, steering, and Bell nonlocality, are evaluated for three coupled qubits in the steady-state configuration. We employ the phenomenological master equation and the microscopic master equation to probe such quantum resources, which provide very different results depending on the system configuration. In particular, steering and Bell nonlocality are null within the phenomenological model, while they reach considerable values within the microscopic model. These results show that the phenomenological approach is not able to capture all quantum resources of the system. We also provide an analytical expression for the steady-state and quantum resources of the system composed of three coupled qubits in the zero temperature limit. Such results demonstrate that quantum resources between two qubits are strongly affected by the third qubit in a nontrivial way.
\end{abstract}


\end{frontmatter}


\section{Introduction}

Quantum resources play a fundamental role in new technologies, demonstrating an advantage over classical resources in computational and informational processing protocols. One of the most known quantum resources is entanglement~\cite{horodecki2009}, which can be used for tasks such as quantum cryptography~\cite{gisin2002}, randomness generation~\cite{pironio2010}, and quantum metrology~\cite{toth2014}. Another important quantum resource is Bell nonlocality~\cite{brunner2014}, identified as a resource for tasks in quantum communication~\cite{buhrman2010} and cryptography~\cite{ekert1991}. Most recently, another resource has been identified and it is known as quantum steering~\cite{uola2020}, and it is useful for one-sided device-independent processing in quantum key distribution~\cite{branciard2012} and subchannel discrimination~\cite{piani2015}. 
Although these resources represent different aspects of quantum correlations, in the case of pure states, they are all equivalent. However, for mixed states, they respect a hierarchy relation, i.e. not every entangled state is steerable and not every steerable state is Bell nonlocal, and steering lies in between entanglement and Bell nonlocality~\cite{quintino2015}. 

Quantum resources generally fade out due to decoherence effects, but there are some schemes where decoherence can be beneficial. In such situations, the steady-state of an open quantum system exhibit the desired quantum resource~\cite{arnese2001,hartmann2006,huelga2012,sauer2014,huang2018}. In Ref.~\cite{fischbach2015}, the entanglement between two qubits $A$ and $B$ is enhanced when a third qubit $C$ is symmetrically coupled to both $A$ and $B$ by an Ising interaction. Such analysis was performed in the steady-state limit, which was numerically obtained by solving the phenomenological master equation (PME)~\cite{breuer2007}. Several investigations have been carried out employing the PME to probe the dynamics of qubits coupled to an environment, {\it e.g.} Refs.~\cite{cattaneo2019,barranco2017} and references therein. The PME describes the nonunitary part of the dissipative dynamics by the sum of terms related to the individual dissipation channel. Such a description of decoherence may lead to controversial physical results~\cite{cattaneo2019,werlang14,werlang15,werlang2020,bellomo2020}. To correctly describe the dynamics of interacting systems, the microscopic master equation (MME)~\cite{breuer2007,rivas2012} must be used. Within this approach, the nonunitary part of the dissipative dynamics is described by global effects, leading to a correct physical description of the nonequilibrium dynamics in open quantum systems.

In this paper, we compare the behavior of different quantum resources using both models, the PME and the MME. Although some investigations have been carried out recently in the context of nonequilibrium quantum resources employing the MME~\cite{wang2020arxiv}, further discussions are still lacking in the literature. We also want to address the question about the influence of a third qubit on the quantum resources obtained from a system of two coupled qubits interacting with this third qubit. Moreover, we provide analytical expressions for the quantum resources within the MME approach showing that the dependence on the system and coupling parameters is quite complicated, but quantum resources are strongly affected due to the presence of the third qubit.

For some specific tasks in quantum information processing, the presence of entanglement is not enough, and it is also necessary the presence of Bell nonlocality and/or quantum steering. Here, we want to show whether these resources can be detected in the steady-state using the PME and the MME models. Interestingly, we show that these resources are only detected in the description using the MME model. Furthermore, we show that the effect of enhancement due to the coupling of two qubits $A$ and $B$ to a third qubit $C$, reported for entanglement in~\cite{fischbach2015}, is also present for quantum steering and Bell nonlocality. We also show that the entanglement between qubits $A$ and $B$ exhibit a nontrivial behavior as a function of the coupling magnitude to the third part. All these results demonstrate that the MME model must be employed to correctly obtain the quantum resources and that the coupling of two qubits to a third part reveals a counterintuitive behavior as a function of the system parameters.

This paper is organized as follows. In Sec.~\ref{model}, we detail the model we investigate in this work, which is composed of two interacting qubits coupled to a third system. In Sec.~\ref{diss-dyn}, we describe the dissipative dynamics, governed by the PME and the MME. We introduce the measures of the quantum resources (entanglement, quantum steering, and Bell nonlocality) in Sec.~\ref{resources}. We reserve the Sec.~\ref{results} for the description of the results, and Sec.~\ref{conclusions} for conclusions. Analytical results for the steady-state within the MME model are described in Appendix A.

\section{Model}
\label{model}
Consider a model with the following Hamiltonian~\cite{fischbach2015}
\begin{equation}
H_S = H_{f}+H_{int},
\label{HFull}
\end{equation}
where
\begin{equation}
H _{f} = \frac{\hbar\omega_{0}}{2}\left ( \sigma_{z}^A + \sigma_{z}^B\right )+\frac{\hbar\omega_{C}}{2}\sigma_{z}^C, \label{H0array}
\end{equation}
is the free Hamiltonian of the system, composed by the three qubits $A$, $B$, and $C$, and $\sigma_i$, with $i=\{x,y,z\}$, are the Pauli matrices. Here, we assume that qubits $A$ and $B$ are identical; thereby, having the same eigenfrequency $\omega_A=\omega_B=\omega_{0}$. The eigenfrequency $\omega_C$ is related to the third qubit $C$. The  Hamiltonian that describes the interaction between the qubits is
\begin{equation}
H _{int} = \hbar J\sigma_{x}^A\sigma_{x}^B + \hbar J_C\left ( \sigma_{x}^A\sigma_{x}^C + \sigma_{x}^B\sigma_{x}^C \right ),
\label{Hint-array}
\end{equation}
where $J$ is the coupling strength between qubits $A$ and $B$, and $J_C$ is the coupling strength between the identical qubits and qubit $C$. For simplicity, we assume $\hbar$ = 1, and all physical parameters are measured in units of the inverse of time, which can be related to the specific physical implementation of qubits, {\it e. g.} solid-state qubits with frequencies in the range of gigahertz~\cite{blais2021}.

\section{Dissipative dynamics}
\label{diss-dyn}
In Ref.~\cite{fischbach2015}, the authors show that the qubits $A$, $B$, and $C$ suffer the influence of thermal baths at zero-temperature, which can be described by a coupling of the individual systems to a continuum of harmonic oscillators~\cite{breuer2007}. By assuming the weak coupling interaction between system-environment and the Born, Markov, and secular approximation~\cite{breuer2007,rivas2012}, we arrive at the PME
\begin{equation}
\label{eq:PMEspin}
\dot{\rho}_{S}(t) = \mathcal{L}[\rho_{S}(t)] = -i[H_{S},\rho_{S}(t)] + \mathcal{D}\left[\rho_{S}(t)\right],
\end{equation}
where $\rho_S(t)$ is the time-dependent density matrix of the system composed of three qubits and $\mathcal{L}$ is the Liouvillian \cite{Albert}. The phenomenological dissipator is given by
\begin{equation}
\label{eq:DFspin}
\mathcal{D}\left[\rho_{S}(t)\right] = \sum_{j=A,B,C}\left[\mathcal{O}_{j}\rho_{S} \mathcal{O}_{j}^{\dagger} -
\frac{1}{2}\left\{\mathcal{O}_{j}^{\dagger }\mathcal{O}_{j},\rho _{S}\right\}\right ],
\end{equation}
with $ \mathcal{O}_{j} = \sqrt{\gamma_{j}}\sigma^{j}_{-}$. The decoherence parameters are given by $\gamma_j$ and $\sigma^{j}_{-}=(\sigma^{j}_{x}-i\sigma^{j}_{y})/2$. The MME has the same form as Eq.~(\ref{eq:PMEspin}), but the dissipator is different; therefore, Eq.~(\ref{eq:DFspin}) should be replaced by the following form
\begin{eqnarray}
\label{eq:Dmic}
\mathcal{D}\left[\rho_{S}(t) \right]=\sum_{j=A,B,C}\sum_{\omega }J\left(
\omega \right) \left[ A_{j}(\omega )\rho _{S}A_{j}^{\dagger }(\omega )-\frac{1}{2}\left\{ A_{j}^{\dagger }(\omega )A_{j}(\omega ),\rho _{S}\right\}
\right],
\end{eqnarray}
where $J(\omega )=\gamma_j\omega$ is the Ohmic spectral density and the operator
\begin{eqnarray}
\label{Ai}
A_{j}(\omega) = \sum_{\epsilon_k-\epsilon_i=\omega}|\epsilon _i\rangle\langle\epsilon _i| A_{j}|\epsilon _k\rangle\langle \epsilon_k|,
\end{eqnarray}
describes the coupling of the system with the environment, where $\omega$ is the energy difference relating the eigenstates $|\epsilon _i\rangle$ and $|\epsilon _k\rangle$ of the full Hamiltonian (Eq.~(\ref{HFull})), thus $H_S|\epsilon _i\rangle=\epsilon _i|\epsilon _i\rangle$. For the heat bath, we have spontaneous decay which can be described by using $A_{j} = \sigma_x^{j}$.

It is well known that one possible steady-state solution of the MME is the Gibbs state $\rho_S(\infty)=\exp\left(-\beta H_S \right)/\textrm{Tr}\{\exp\left(-\beta H_S \right)\}$~\cite{werlang2020}, where  $\beta=1/(k_BT)$, with $k_B$ being the Boltzmann's constant and $T$ being the bath temperature. On the other hand, the dissipation operators of Eq.~(\ref{eq:PMEspin}) act locally in each subsystem, therefore leading to a steady-state solution different to the Gibbs state. 
The main focus of this paper is to explore the difference in the steady-state solution when working with the phenomenological or microscopic models in the quantum resources of three qubits $A$, $B$, and $C$.

\section{Quantum resources}
\label{resources}

In this work, we analyze the behavior of three different quantum resources, entanglement, steering and Bell nonlocality. In the following, we define appropriate measures for each one of these quantum resources.

Entanglement is a well-known quantum resource, which reflects the nonseparability of a quantum state~\cite{horodecki2009}. In the case of mixed two-qubit states, several measures have been proposed, but here we focus on a measure known as concurrence~\cite{wootters1998}, given by,
\begin{equation}
E\left(\rho\right) =\max \Big\{0,\sqrt{\lambda_{1}} - \sqrt{\lambda_{2}} - \sqrt{\lambda_{3}} - \sqrt{\lambda_{4}}\Big\}, \label{Conc}
\end{equation}%
where $\lambda_i$ are the eigenvalues of the matrix $R = \tilde{\rho}\rho $, with $\tilde{\rho}\equiv\left(
\sigma_{y}\otimes \sigma_{y}\right) \rho^{\ast}\left(\sigma_{y}\otimes \sigma _{y}\right)$, and $\rho^{\ast}$ is the complex conjugate of $\rho$.

Steering, a term coined by Schr\"odinger~\cite{schrodinger1935,schrodinger1936} in a reply to the Einstein-Podolsky-Rosen (EPR) argument of the  incompleteness  of quantum mechanics, is the quantum resource that name the ability of an observer to disturb (or steer) another state in a remote site through local measurements~\cite{wiseman2007,uola2020}.
One way of detecting steering is through an inequality that is composed of a finite sum of expectation values of $n=2,3$ observables~\cite{cavalcanti2009}. A two-qubit system is nonsteerable if
\begin{equation}
F_n(\rho,\mu)=\frac{1}{\sqrt{n}}\left|\sum_{i=1}^n \langle A_i\otimes B_i\rangle\right| \leqslant 1, \label{steering}
\end{equation}
where $A_i = \hat{u}_i\cdot\vec{\sigma}$, $B_i = \hat{v}_i\cdot\vec{\sigma}$, $\vec{\sigma} = (\vec{\sigma}_1,\vec{\sigma}_2,\vec{\sigma}_3)$ is a vector composed of the Pauli matrices, $\hat{u}_i \in\mathbb{R}^3$ are unit vectors, $\hat{v}_i \in\mathbb{R}^3$ are orthonormal vectors, $\mu = \{\hat{u}_1, \cdots, \hat{u}_n, \hat{v}_1, \cdots, \hat{v}_n\}$ is the set of measurement directions, and $\langle A_i\otimes B_i\rangle = \textrm{Tr}[\rho (A_i\otimes B_i)]$. Here we focus on the three measurement criterion, since it has been shown that the one based on two measurements is a nonlocality measure~\cite{costa2016}.

A measure of steering in the three-measurement scenario based in the above inequality was proposed in~\cite{costa2016}, and is given by
\begin{equation}
S(\rho) = \max \left\{0, \frac{F_3(\rho)-1}{F_3^{\max}-1}\right\},
\end{equation}%
where $F_3(\rho) = \max_{\mu} F_3(\rho,\mu)$ and $F_3^{\max} = \max_{\rho}F_3(\rho) = \sqrt{3}$, with $S\in [0,1]$.

If the correlations of a quantum state cannot be explained in terms of any local hidden variable model (LHVM), it is said that this state is Bell nonlocal, and it can be detected by violations of Bell-like inequalities. In the case of two-qubit systems, the Clauser-Horne-Shimony-Holt (CHSH) inequality~\cite{clauser1969} is usually employed, and can be written as
\begin{equation}
B(\rho,\mu) = |\textrm{Tr}(\rho B_{CHSH})| \leqslant 2,
\end{equation}
where $B_{CHSH}$ is a Bell operator that involves Alice and Bob's measurement settings $\mu$. A measure of Bell nonlocality, based in the Refs.~\cite{horodecki1995,hu2013,costa2016}, can be given by $B(\rho):=\max_{\mu} B(\rho,\mu)$. A measure of Bell nonlocality, up to a suitable normalization, can be written as~\cite{costa2016a}
\begin{equation}
N(\rho) = \max\left\{0, \frac{B(\rho)-2}{B_{\max}(\rho)-2}\right\},
\end{equation}%
where $B_{\max}(\rho) = 2\sqrt{2}$, which is known as the Tsirelson's bound.

As shown in~\cite{costa2016,horodecki1995}, steering and Bell nonlocality based in the above measures have closed expressions, if one expresses the two-qubit states in its Bloch representation form~\cite{luo2008},
\begin{equation}
\rho = \frac{1}{4}\left(\textbf{1}\otimes\textbf{1} + \vec{a}\cdot\vec{\sigma}\otimes\textbf{1} + \textbf{1}\otimes\vec{b}\cdot\vec{\sigma} + \sum_{i=1}^3 c_i\sigma_i\otimes\sigma_i\right),
\end{equation}%
where $\textbf{1}$ is the $2\times 2$ identity matrix and $\{\vec{a},\vec{b},\vec{c}\}\in\mathbb{R}^3$ are vectors with norm less than unit. In this representation, steering can be evaluated using the following closed formula,
\begin{equation}
\label{steer}
S(\rho) = \max\left\{0, \frac{c-1}{\sqrt{3}-1}\right\},
\end{equation}%
and Bell nonlocality,
\begin{equation}
\label{bell}
N(\rho) = \max\{0, \frac{\sqrt{c^2 - c^2_{\min}}-1}{\sqrt{2}-1}\},
\end{equation}%
where $c = \sqrt{c_1^2+c_2^2+c_3^2}$ and $c_{\min}\equiv \min\{|c_1|,|c_2|,|c_3|\}$.

By the use of the quantifiers given in Eqs.~(\ref{Conc}),~(\ref{steer}), and~(\ref{bell}), we are able to analyze the behavior of the quantum resources described above in the model we are considering in this work. Note that these measures are defined in order to be contained in the interval $[0,1]$, with 0 corresponding to the case of absence of the quantum resource, and 1 to its maximum amount.

\section{Results}
\label{results}
\begin{figure}
\begin{center}
\includegraphics[scale=0.8,angle=0]{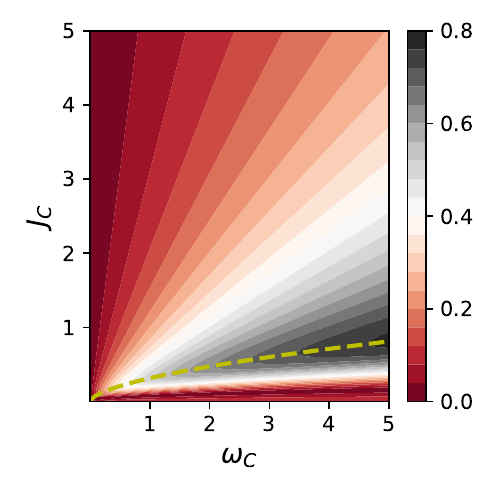}
\end{center}
\caption{Entanglement $E\left(\rho^{MME}_{A,B}\right)$ evaluated for the steady-state obtained from the MME as a function of the parameters of the third qubit, $J_{C}$ and $\omega_C$, in the case of a weak coupling ($J=0.01$) between qubits $A$ and $B$. Entanglement is maximum for $J_C\approx 0.31\omega_C^{0.58}$ (dashed curve). Here, $\omega_{0}=0.1$.}
\label{fig1}
\end{figure}

\begin{figure}
\begin{center}
\includegraphics[scale=0.8,angle=0]{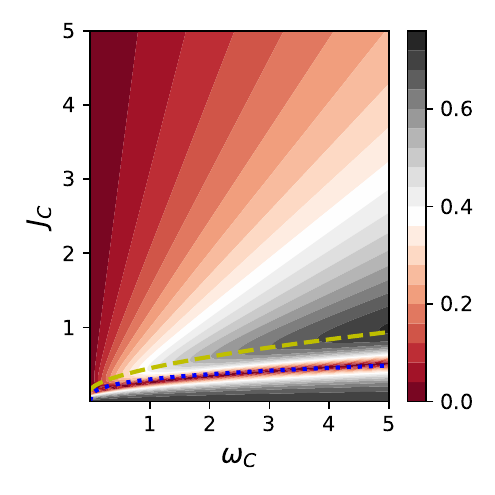}
\end{center}
\caption{ Entanglement $E\left(\rho^{MME}_{A,B}\right)$ evaluated for the steady-state obtained from the MME as a function of the parameters of the third qubit, $J_{C}$ and $\omega_C$, in the case of a coupling ($J=0.1$) between qubits $A$ and $B$. Entanglement is minimum for $J_C\approx 0.3\omega_C^{0.3}$ (dotted curve) and is maximum for $J_C\approx 0.15+0.3\omega_C^{0.6}$ (dashed curve). Here, $\omega_{0}=0.1$.}
\label{fig2}
\end{figure}

\begin{figure}[ht!]
\begin{center}
\includegraphics[scale=0.8,angle=0]{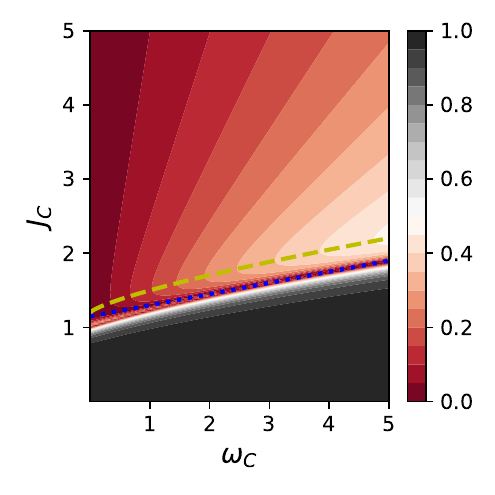}
\end{center}
\caption{Entanglement $E\left(\rho^{MME}_{A,B}\right)$ evaluated for the steady-state obtained from the MME as a function of the parameters of the third qubit, $J_{C}$ and $\omega_C$, in the case of a strong coupling ($J=1$) between qubits $A$ and $B$. Entanglement is minimum for $J_C\approx 1.15+0.15\omega_C$ (dotted curve) and maximum for $J_C\approx 1.2+0.3\omega_C^{0.75}$ (dashed curve). Here, $\omega_{0}=0.1$.}
\label{fig3}
\end{figure}

We start our analysis by considering the MME and by calculating the entanglement for the two qubits $A$ and $B$, using the reduced density matrix $\rho^{MME}_{A,B} = \textrm{Tr}_{C}\left\{\rho_{SS}\right\}$. The symbol $\textrm{Tr}_{C}$ indicates the partial trace over the qubit $C$. In Eq.~(\ref{conc_mme}) of Appendix A, we provide an analytical form for the entanglement $E\left(\rho^{MME}_{A,B}\right)$ within the MME approach, showing that it depends on $\omega_{0}$, $\omega_C$, $J$, and $J_C$ in a complicated non-monotonic form. To extract some insights about this complicated multivariate function, we fix some parameters and vary others in Figs.~\ref{fig1}-\ref{fig3}.

In Fig.~\ref{fig1}, we plot the entanglement $ E\left(\rho^{MME}_{A,B}\right)$ as a function of the eigenfrequency and the coupling strength related to the qubit $C$ considering  $\omega_{0}=0.1$ and $J=0.01$. The entanglement is small for $\omega_C<1$ and has a maximum value for $\omega_C=5$ and $J_C\approx 0.8$. In this case, the direct coupling between qubits $A$ and $B$ is small and the entanglement between them only occurs by their coupling to a third part (qubit $C$). Thus, it is expected that the entanglement increases as $J_C$ rises. On the other hand, the dependence on $\omega_C$ is not clear in Eq.~(\ref{conc_mme}), but we find that the maximum entanglement occurs when $J_C\approx 0.31\omega_C^{0.58}$, which is shown by the dashed curve in Fig.~\ref{fig1}.
In Fig.~\ref{fig2}, we provide a similar contour plot as shown in Fig.~\ref{fig1}, but setting the coupling between qubits $A$ and $B$ equals to the value of their eigenfrequencies $J= \omega_{0}=0.1$. One can see that the entanglement has an appreciable value for small $J_C$ because the coupling between qubits $A$ and $B$ is capable of creating entanglement between them in this case. When $J_C$ surpass the value of $J$, qubits $A$ and $B$ start to couple to qubit C and $E\left(\rho^{MME}_{A,B}\right)$ goes almost to zero for $J_C\approx 0.3\omega_C^{0.3}$ (dotted curve in Fig.~\ref{fig2}). By further increasing $J_C$, a maximum value is found for $J_C\approx 0.15+0.3\omega_C^{0.6}$ (dashed curve in Fig.~\ref{fig2}).

In the contour plot shown in Fig.~\ref{fig3}, we plot the entanglement for $\omega_{0}=0.1$ and $J=1$. In this case, the entanglement between qubits $A$ and $B$ is close to 1 when $J_C<1$ due to the Ising interaction in Eq.~(3). When $J_C$ approaches to $J=1$,  the qubit $A$ will interact to both qubits $B$ and $C$ with similar amplitudes; thus, the entanglement between qubits $A$ and $B$ drastically decays reaching a minimum value for $J_C\approx 1.15+0.15\omega_C$ (dotted curve in Fig.~\ref{fig3}). By further increasing $J_C$, $E\left(\rho^{MME}_{A,B}\right)$ reaches a maximum value when $J_C\approx 1.2+0.3\omega_C^{0.75}$ (dashed curve in Fig.~\ref{fig3}). Such results presented in Figs.~\ref{fig1}-\ref{fig3} demonstrate how the steady-state of these three qubits depends on system parameters in a nontrivial way; thereby, obscuring the physical intuition about the entanglement between qubits $A$ and $B$ in the system described by Eqs.~(2) and (3), which reinforce the necessity of numerical simulations.

In order to compare the MME and the PME models, we first consider the case where $J_C=0$. For the PME model, entanglement does not depend on $\omega_C$ and it is given by the analytical formula found in Ref.~\cite{sauer2014}, which reads
\begin{equation}
 E\left(\rho^{PME}_{A,B}\right)=\max\left\{0,\frac{|x|-1/2}{1+|x|^2}\right\},\label{conc_pme_jc0}
\end{equation}
where $x=(\omega_{0}+i\gamma)/J$. In Eq.~(\ref{conc_pme_jc0}), we can see that the entanglement is almost zero when $J>>\max\{\omega_{0},\gamma\}$ or $J<<\max\{\omega_{0},\gamma\}$. The maximum entanglement is $E\left(\rho^{PME}_{A,B}\right)\approx 0.31$ when $\sqrt{\omega_{0}^2+\gamma^2}/J=(1+\sqrt{5})/2$~\cite{sauer2014}. When dealing with the MME and $J_C=0$, we can also find an analytical solution for the entanglement, which is given by
\begin{equation}
 E\left(\rho^{MME}_{A,B}\right)=2\left|\frac{J}{\left(\omega_{0}-\sqrt{\omega_{0}^2+J^2}\right)\left[1+\frac{J^2}{\left(\omega_{0}-\sqrt{\omega_{0}^2+J^2}\right)^2}\right]}\right|.
\label{conc_mme_jc0}
\end{equation}
In Eq.~(\ref{conc_mme_jc0}), the entanglement is also almost zero when $J<<\omega_{0}$, but it tends to one when $J>>\omega_{0}$. By comparing these two limits in Eqs.~(\ref{conc_pme_jc0}) and~(\ref{conc_mme_jc0}), we already see that the quantum resources of the steady-state evaluated either by the PME or by the MME models can be very different.

\begin{figure}[ht!]
\begin{center}
\includegraphics[scale=1,angle=0]{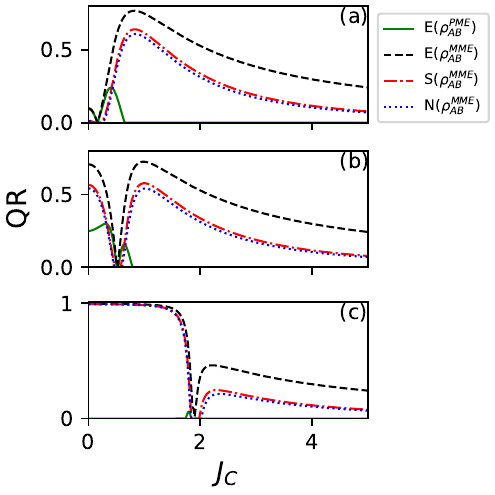}
\end{center}
\caption{Entanglement $E\left(\rho^{MME}_{A,B}\right)$ (dashed black curve), quantum steering $S\left(\rho^{MME}_{A,B}\right)$ (dash-dotted red curve) and Bell nonlocality $S\left(\rho^{MME}_{A,B}\right)$ (dotted blue curve) evaluated for the steady-state obtained from the MME, and $E\left(\rho^{PME}_{A,B}\right)$ (solid green curve) from the PME as a function of the coupling to the third qubit $J_{C}$. Since quantum steering and Bell nonlocality from the PME are both zero, they do not appear in the plot. Here, we vary the coupling between the qubits and the ancilla as (a)  $J=0.01$, (b)  $J=0.1$, and (c)  $J=1.0$. Other parameters are fixed as follows: $\omega_{0}=0.1$ and $\omega_C=5$.}
 \label{fig4}
\end{figure}

In Fig.~\ref{fig4}, we plot the entanglement  calculated through the MME (solid curves) and the PME (dashed curves) models as a function of the coupling strength $J_C$ for a fixed value of the eigenfequency $\omega_C$. In panel (a) of Fig.~\ref{fig4}, we consider $J=0.01$ and $\omega_{0}=0.1$. As expected, the entanglement is small and both models, the PME and the MME, are in agreement for $J_C<0.3$. The entanglement evaluated using the PME model reaches a maximum of $ E\left(\rho^{PME}_{A,B}\right)\approx0.24$ for $J_C\approx0.4$ and decays to zero for $J_C>0.65$. On the other hand, the MME model provides a maximum value of entanglement
$E\left(\rho^{MME}_{A,B}\right)\approx0.78$ for $J_C\approx0.8$ and it slowly decays as $J_C$ increases. Another interesting aspect observed in panel~(a) of Fig.~\ref{fig4} is the fact the entanglement evaluated from the MME is highly enhanced by the presence of the third qubit, going from $E\left(\rho^{MME}_{A,B}\right)\approx0.1$ for $J_C=0$ to $E\left(\rho^{MME}_{A,B}\right)\approx0.78$ for $J_C\approx0.8$. In the middle panel of Fig.~\ref{fig4}, we consider $J=0.1$ and $\omega_{0}=0.1$. For $J_C=0$, solutions coming from the PME and the MME give a different result for the entanglement, whereas the PME gives $E\left(\rho^{PME}_{A,B}\right)\approx0.25$, the MME gives $E\left(\rho^{MME}_{A,B}\right)\approx0.7$. This result shows that even for two qubits with a reasonable coupling, the steady-state is very different. By increasing $J_C$, entanglement evaluated from the PME and the MME achieve a minimum  at $J_C\approx0.5$. The PME model provides another peak at $J_C\approx0.64$ and becomes zero for $J_C>0.8$, while the MME model gives a maximum value for entanglement at $J_C\approx0.92$, and this quantum correlation between the qubits remains up to $J_C=5$. The bottom panel of Fig.~\ref{fig4} shows the entanglement assuming $J=1$ and $\omega_{0}=0.1$. The steady-state evaluated from the PME provides an almost null entanglement and only is different from zero for $J_C=1.8$. In contrast, the entanglement obtained from the MME is almost 1 for $J_C<1$, reaches its minimum value for $J_C\approx1.9$, rapidly increases to 0.45 as $J_C$ increases, and slowly decays as a function of $J_C$.

In Fig.~\ref{fig4}, we also plot the quantum steering (dash-dotted curve) and Bell nonlocality (dotted curve) for the steady-state obtained by the MME. One can notice that steering and Bell nonlocality have similar behavior to entanglement as a function of $J_C$, and, as expected from the hierarchic relation among them, they are nonzero only when the system is entangled. 
On the other hand, steering and Bell nonlocality for the steady-state obtained by the PME are both zeros because $c<1$ and $c^2-c_{\min}<1$.  This result highlights even more the difference between the MME and the PME. While in the MME model the inequalities for steering and Bell nonlocality are violated, the same does not occur for the PME model. 
Lastly, we can see in Fig.~{4} that in the limit where $J_C\rightarrow \infty$, the quantum resources between systems $A$ and $B$ go to zero because the interaction with the qubit $C$ becomes very strong and only the quantum resources between systems $AC$ or $BC$ could be extracted in this limit.

Finally, we mention a setup where our investigation could be experimentally verified. In Ref.~\cite{niskanenPRB}, three  superconducting flux qubits were coupled between each other, with coupling between neighboring qubits reaching values up to 6.05 GHz. In addition, the eigenfrequencies of these qubits can be accordingly manipulated to fit our theoretical predictions.

\section{Conclusions}
\label{conclusions}

In this work, we investigated the influence of a third qubit on the entanglement, steering, and Bell nonlocality, of two coupled qubits at zero temperature in steady-state configuration. We found that quantum resources depend on the coupling of the two qubits Aand Bwith the qubit Cin a nontrivial way. We also showed that the steady-state of the system evaluated from the PME and the MME lead to a completely different set of results for the quantum resources and only the MME model is able to capture the quantum resources properly. Specifically, we show that steering and Bell nonlocality are zero when the PME is employed, which is in contrast to the MME approach that exhibit a finite amount of steering and Bell nonlocality. Such results demonstrate that the MME model must be used to correctly extract the quantum information of interacting systems.

\section{Acknowledgements}
The authors would like to thank T. Werlang for fruitful discussions about this topic.
A.C.S.C. acknowledges support by CAPES/Brazil.
L.K.C. thanks to the Brazilian Agencies FAPESP (grant No 2019/09624-3) and CNPq (grant 311450/2019-9) for supporting this research.

\bibliography{Diniz2021}

\section{Appendix A: Microscopic master equation approach: steady-state}
\label{two-qubits}

The eigenvalues and eigenstates of the ground state for the three qubits can be analytically determined and they are given by,  respectively,
\begin{eqnarray}
\lambda_G &=& \frac{1}{6}\left(2J - \omega_C - \frac{(1 + i\sqrt{3}) \alpha}{\Big(\beta + \sqrt{\beta^2 - \alpha^3}\Big)^{1/3}} - (1 - i\sqrt{3})\Big(\beta + \sqrt{\beta^2 - \alpha^3}\Big)^{1/3}\right),
\nonumber \\
|\psi_G\rangle &=& \frac{1}{\sqrt{\mathcal{N}}}
\left(
\begin{array}{c}
0 \\
-\frac{\gamma + 4\lambda_G(-J+\omega_{0}+\lambda_G)}{2(\delta - 2J \lambda_G)} \\
-\frac{\tau + 2J_C \lambda_G}{\delta - 2J \lambda_G} \\
0 \\
-\frac{\tau + 2J_C \lambda_G}{\delta - 2J \lambda_G} \\
0 \\
0 \\
1
\end{array}
\right), \nonumber
\end{eqnarray}
with normalization
\begin{eqnarray}
\mathcal{N} &=& 1 + 2\left|\frac{\tau + 2J_C \lambda_G}{\delta - 2J \lambda_G}\right|^2 + \left|\frac{\gamma + 4\lambda_G(-J + \omega_{0} + \lambda_G)}{2(\delta - 2J\lambda_G)}\right|^2, \nonumber
\end{eqnarray}
and
\begin{eqnarray}
\alpha &=& 4J^2 + 12 J_C^2 + 3\omega_{0}^2 + \omega_C^2 + 2J\omega_C, \nonumber \\
\beta &=& -8J^3 + 72J J_C^2 - 9J\omega_{0}^2 + 3J\omega_C^2  - 6J^2\omega_C + 18J_C^2 \omega_C \nonumber \\
&& - 9\omega_{0}^2\omega_C + \omega_C^3, \nonumber \\
\gamma &=& -8J_C^2 - (2J+\omega_C)(2\omega_{0}+\omega_C), \nonumber \\
\delta &=& 2J^2 - 4J_C^2 + J\omega_C, \nonumber \\
\tau &=& J_C(J + 2\omega_{0} + \omega_C). \nonumber
\end{eqnarray}

From the ground state given above, we are able to construct the density operator for the qubits $A$ and $B$ after tracing out the ancilla, i.e., $\rho_{A,B}^{MME} = \textrm{Tr}_C |\psi_G\rangle\langle\psi_G|$, 
\begin{eqnarray}\label{rhoab}
\rho_{A,B}^{MME} &=& \frac{1}{\mathcal{N}}
\left(
\begin{array}{cccc}
\left|\frac{\gamma - 4(J-\omega_{0})\lambda_G+ 4\lambda_G^2}{2(\delta - 2J \lambda_G)}\right|^2 & 0 & 0 & -\frac{(\gamma - 4(J-\omega_{0})\lambda_G+4\lambda_G^2)}{2(\delta - 2J\lambda_G)} \\
0 & \left|\frac{\tau + 2J_C\lambda_G}{\delta - 2J\lambda_G}\right|^2 & \left|\frac{\tau + 2J_C\lambda_G}{\delta - 2J\lambda_G}\right|^2 & 0 \\
0 & \left|\frac{\tau + 2J_C\lambda_G}{\delta - 2J\lambda_G}\right|^2 & \left|\frac{\tau + 2J_C\lambda_G}{\delta - 2J\lambda_G}\right|^2 & 0 \\
-\frac{(\gamma - 4(J-\omega_{0})\lambda_G^\ast+\lambda_G^{\ast^2})}{2(\delta - 2J\lambda_G^\ast)} & 0 & 0 & 1
\end{array}
\right). \nonumber
\end{eqnarray}
For the state $\rho^{MME}_{A,B}$, entanglement is given by
\begin{equation}\label{conc_mme}
E(\rho^{MME}_{A,B}) = 2\max\left\{0, \frac{1}{\mathcal{N}} \left|\left|\frac{\gamma - 4(J-\omega_{0})\lambda_G+4\lambda_G^2}{2(\delta - 2J\lambda_G)}\right| - \left|\frac{\tau + 2J_C\lambda_G}{\delta - 2J\lambda_G}\right|^2\right|\right\}, 
\end{equation}
and steering~(\ref{steer}) and Bell nonlocality~(\ref{bell}) can be evaluated from
\begin{eqnarray}
c_1 &=& -\frac{2}{\mathcal{N}}\left(\left|\frac{\gamma - 4(J-\omega_{0})\lambda_G+4\lambda_G^2}{2(\delta - 2J\lambda_G)}\right| + \left|\frac{\tau + 2J_C\lambda_G}{\delta - 2J\lambda_G}\right|^2\right), \nonumber \\
c_2 &=& -\frac{1}{\mathcal{N}}\left(1+\left|\frac{\gamma - 4(J-\omega_{0})\lambda_G+4\lambda_G^2}{2(\delta - 2J\lambda_G)}\right|^2 -2 \left|\frac{\tau + 2J_C\lambda_G}{\delta - 2J\lambda_G}\right|^2\right), \nonumber \\
c_3 &=& -\frac{2}{\mathcal{N}}\left(\left|\frac{\gamma - 4(J-\omega_{0})\lambda_G+4\lambda_G^2}{2(\delta - 2J\lambda_G)}\right| - \left|\frac{\tau + 2J_C\lambda_G}{\delta - 2J\lambda_G}\right|^2\right). \nonumber
\end{eqnarray}

\end{document}